\documentclass[aps,prl, reprint, superscriptaddress, floatfix, amssymb, amsfonts, showpacs]{revtex4-1}
\usepackage{epsf,amsmath,amssymb,verbatim,color,multirow,pifont}
\usepackage{graphicx}
\pdfoutput=1

\begin{document}
\title{Unconventional anomalous Hall effect from antiferromagnetic domain walls of $\text{Nd}_{\text{2}}$$\text{Ir}_{\text{2}}$$\text{O}_{\text{7}}$ thin film
}

\author{Woo Jin Kim}
\affiliation{Center for Correlated Electron Systems, Institute for Basic Science, Seoul 08826, Republic of Korea}
\affiliation{Department of Physics and Astronomy, Seoul National University, Seoul 08826, Republic of Korea}
\author{John H. Gruenewald}
\affiliation{Department of Physics and Astronomy, University of Kentucky, Lexington, KY 40506, USA}
\author{Taekoo Oh}
\affiliation{Center for Correlated Electron Systems, Institute for Basic Science, Seoul 08826, Republic of Korea}
\affiliation{Department of Physics and Astronomy, Seoul National University, Seoul 08826, Republic of Korea}
\affiliation{Center for Theoretical Physics (CTP), Seoul National University, Seoul 08826, Republic of Korea}
\author{Sangmo Cheon}
\affiliation{Department of Physics, Hanyang University, Seoul 04763, Republic of Korea}
\author{Bongju Kim}
\affiliation{Center for Correlated Electron Systems, Institute for Basic Science, Seoul 08826, Republic of Korea}
\affiliation{Department of Physics and Astronomy, Seoul National University, Seoul 08826, Republic of Korea}
\author{Oleksandr B. Korneta}
\affiliation{Center for Correlated Electron Systems, Institute for Basic Science, Seoul 08826, Republic of Korea}
\affiliation{Department of Physics and Astronomy, Seoul National University, Seoul 08826, Republic of Korea}
\author{Hwanbeom Cho}
\affiliation{Center for Correlated Electron Systems, Institute for Basic Science, Seoul 08826, Republic of Korea}
\affiliation{Department of Physics and Astronomy, Seoul National University, Seoul 08826, Republic of Korea}
\author{Daesu Lee}
\affiliation{Center for Correlated Electron Systems, Institute for Basic Science, Seoul 08826, Republic of Korea}
\affiliation{Department of Physics and Astronomy, Seoul National University, Seoul 08826, Republic of Korea}
\author{Yoonkoo Kim}
\affiliation{Department of Materials Science and Engineering and Research Institute of Advanced Materials, Seoul National University, Seoul 08826, Republic of Korea}
\author{Miyoung Kim}
\affiliation{Department of Materials Science and Engineering and Research Institute of Advanced Materials, Seoul National University, Seoul 08826, Republic of Korea}
\author{Je-Geun Park}
\affiliation{Center for Correlated Electron Systems, Institute for Basic Science, Seoul 08826, Republic of Korea}
\affiliation{Department of Physics and Astronomy, Seoul National University, Seoul 08826, Republic of Korea}
\author{Bohm-Jung Yang}
\affiliation{Center for Correlated Electron Systems, Institute for Basic Science, Seoul 08826, Republic of Korea}
\affiliation{Department of Physics and Astronomy, Seoul National University, Seoul 08826, Republic of Korea}
\affiliation{Center for Theoretical Physics (CTP), Seoul National University, Seoul 08826, Republic of Korea}
\author{Ambrose Seo}
\affiliation{Department of Physics and Astronomy, University of Kentucky, Lexington, KY 40506, USA}
\author{Tae Won Noh}
\affiliation{Center for Correlated Electron Systems, Institute for Basic Science, Seoul 08826, Republic of Korea}
\affiliation{Department of Physics and Astronomy, Seoul National University, Seoul 08826, Republic of Korea}
\date{\today}

\begin{abstract}
Ferroic domain walls (DWs) create different symmetries and ordered states compared with those in single-domain bulk materials. In particular, the DWs of an antiferromagnet (AFM) with non-coplanar spin structure have a distinct symmetry that cannot be realized in those of their ferromagnet counterparts. In this paper, we show that an unconventional anomalous Hall effect (AHE) can arise from the DWs of a non-coplanar AFM, $\text{Nd}_{\text{2}}$$\text{Ir}_{\text{2}}$$\text{O}_{\text{7}}$.Bulk $\text{Nd}_{\text{2}}$$\text{Ir}_{\text{2}}$$\text{O}_{\text{7}}$ has a cubic symmetry; thus, its Hall signal should be zero without an applied magnetic field. The DWs generated in this material break the two-fold rotational symmetry, which allows for finite anomalous Hall conductivity. A strong \textit{f}-\textit{d} exchange interaction between the Nd and Ir magnetic moments significantly influences antiferromagnetic domain switching. Our epitaxial $\text{Nd}_{\text{2}}$$\text{Ir}_{\text{2}}$$\text{O}_{\text{7}}$ thin film showed a large enhancement of the AHE signal when the AFM domains switched, indicating that the AHE is mainly due to DWs. Our study highlights the symmetry broken interface of AFM materials as a new means of exploring topological effects and their relevant applications.
\end{abstract}

 \maketitle

\section*{I. INTRODUCTION}

 Both symmetry and the ferroic order inevitably become modified in a ferroic domain wall (DW), triggering new functions and topological properties [1-6]. The discovery of conductive DWs in ferroelectric thin films opened a broad field of research dedicated to understanding the underlying mechanisms, and the fabrication of practical DW-based devices [1-4]. The DWs of certain ferromagnets exhibit topological Hall effects attributable to magnetic skyrmions [5, 6]. However, similar topological properties have yet to be observed in the DWs of antiferromagnets (AFMs). Although there exists a theoretical prediction that a finite anomalous Hall effect (AHE) can arise among AFM DWs with non-vanishing Berry curvatures [7], few experimental studies have investigated this intriguing possibility.

Since Wan \textit{et al.} theoretically proposed the existence of a Weyl fermionic state in AFM-ordered $\text{Y}_{\text{2}}$$\text{Ir}_{\text{2}}$$\text{O}_{\text{7}}$ [8], the rare-earth pyrochlore iridates have attracted great interest [9]. Particularly, $\text{Nd}_{\text{2}}$$\text{Ir}_{\text{2}}$$\text{O}_{\text{7}}$ has been widely studied, due to its closeness to the metal–insulator transition (MIT) [10, 11] and large DW conductance [12]. To discuss the AFM spin structure in detail, consider that $\text{Nd}_{\text{2}}$$\text{Ir}_{\text{2}}$$\text{O}_{\text{7}}$ is composed of two types of tetrahedra, Ir and Nd sublattices, as shown in Fig. 1(a). As shown by the red arrows, all four of the Ir spins at the vertices of a tetrahedron point inward. At the nearest neighboring Ir tetrahedra (not shown), all of the Ir spins should point outward. These results in all-in-all-out (AIAO) ordering of Ir spins, which we refer to as Ir:AIAO. At the same time, the spins in the Nd sublattice have similar magnetic ordering (all-out-all-in), Nd:AOAI, as shown in Fig. 1(a). These intriguing magnetic orderings play important roles in the physical properties of these materials [8-12].

Under an external magnetic field, \textit{H}, we can switch between AIAO and AOAI domains of Ir spins. It is important to understand how DWs can be formed during the switching process in an $\text{Nd}_{\text{2}}$$\text{Ir}_{\text{2}}$$\text{O}_{\text{7}}$ sample. Note that the Nd ion has a larger magnetic moment; thus, the corresponding Zeeman field is larger. When \textit{H} increases along the [111] direction, Nd spins become canted and then suddenly flip to form the 3-in-1-out (3I1O) configuration [13]. At a higher \textit{H}, Nd spins can induce the flip of Ir sublattice spin into the AOAI configuration via the \textit{f}-\textit{d} exchange interaction [see Fig. 1(b)]. During this spin switching process, two kinds of domains may coexist, resulting in the presence of DWs [see Fig. 1(c)].

Here we present an $\text{Nd}_{\text{2}}$$\text{Ir}_{\text{2}}$$\text{O}_{\text{7}}$ epitaxial thin film, as an ideal system for realizing the intriguing topological responses that originate from the DWs of an AFM material. The electronic structure of $\text{Nd}_{\text{2}}$$\text{Ir}_{\text{2}}$$\text{O}_{\text{7}}$ has been assumed to be close to that of a Weyl semimetal [8, 14]. This similarity to a semimetallic ground state results in a very large DW conductance [12, 13]. Within a single AIAO or AOAI domain, AHE is forbidden by cubic crystalline symmetry [15, 16]. However, in the presence of the DW, we will show that two-fold rotation symmetries are explicitly broken, resulting in uncompensated magnetic moments (see Section IV C). This leads to finite anomalous Hall conductivity in the plane parallel to the highly conducting DW. Thus, it will be informative to explore DW-induced unconventional magneto-transport in low-dimensional $\text{Nd}_{\text{2}}$$\text{Ir}_{\text{2}}$$\text{O}_{\text{7}}$ films, in which the contribution of the DW can be maximized [17].

\begin{figure}
\includegraphics{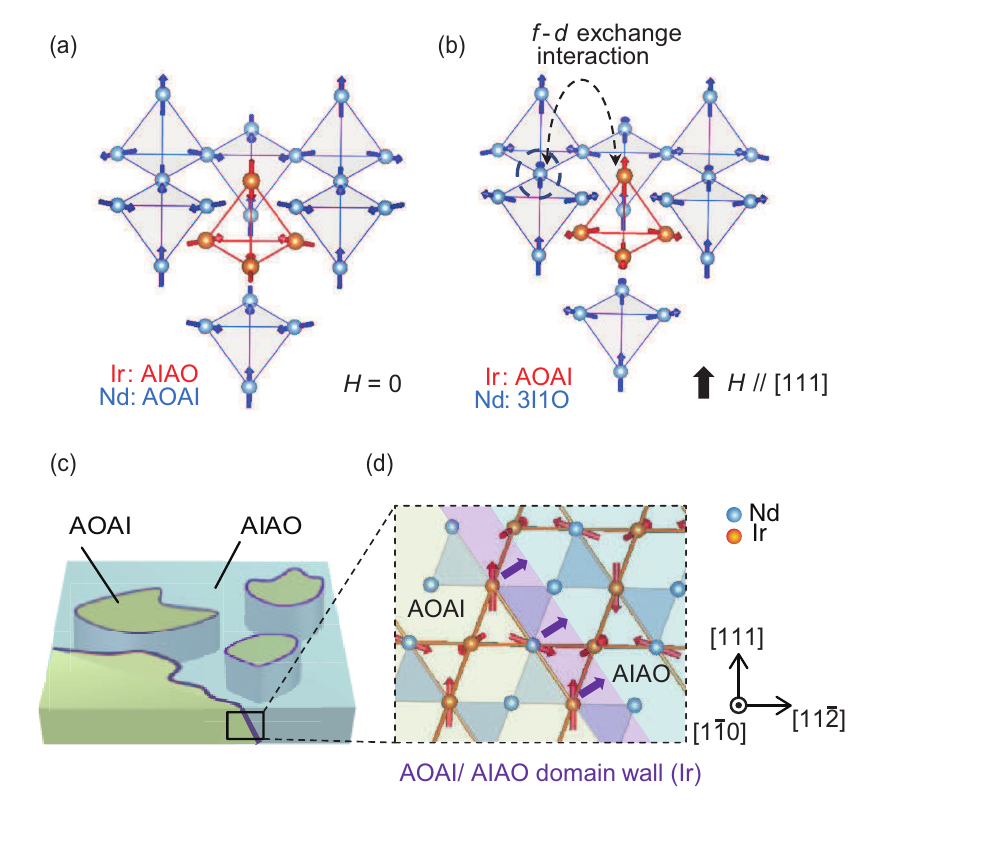}
\caption{(Color online) Schematic diagrams for the magnetic domain structure of $\text{Nd}_{\text{2}}$$\text{Ir}_{\text{2}}$$\text{O}_{\text{7}}$. (a) The magnetic structure of Nd 4\textit{f} moments (blue arrows) and Ir 5\textit{d} moments (red arrows) in \textit{H} = 0 at temperatures below $T_{\text{N}}^{\text{Nd}}$. The Ir and Nd sublattices have AIAO and AOAI ordering, respectively. (b) The magnetic structure in sufficiently strong \textit{H}. The Nd 4\textit{f} and Ir 5\textit{d} moments are coupled with the \textit{f}-\textit{d} exchange interaction. (c) Schematic diagram of the domain walls (DWs) between Ir-AIAO and Ir-AOAI. The purple-colored line indicates the possible domain wall structure with finite magnetization. (d) Schematic diagram of the dotted square region in (c), enlarged for detail. The purple arrows indicate the net magnetization of Ir moments at a DW.
}\label{events}
\end{figure}

\section*{II. EXPERIMENTS}
\subsection{A. Growth of $\text{Nd}_{\text{2}}$$\text{Ir}_{\text{2}}$$\text{O}_{\text{7}}$ films}

We prepared $\text{Nd}_{\text{2}}$$\text{Ir}_{\text{2}}$$\text{O}_{\text{7}}$ (111) films on commercial Y-stabilized $\text{ZrO}_{\text{2}}$ (YSZ) (111) single-crystal substrates via pulsed laser deposition and following post annealing procedure. We irradiated a single-phase $\text{Nd}_{\text{2}}$$\text{Ir}_{\text{2}}$$\text{O}_{\text{7}}$ polycrystalline target with a KrF excimer laser ($\lambda$ = 248 nm); the laser fluence and frequency were 4.5 J/$\text{cm}^{\text{2}}$ and 3 Hz, respectively. We maintained the distance between the target and the substrate at 50 mm. It is well known that the pyrochlore iridate thin films are extremely difficult to grow because of volatility of iridium [18-20]. To form the pyrochlore phase thermodynamically, it is required to use a high oxygen pressure and a high temperature [21]. When we try to grow pyrochlore iridate films in such a condition, a gas phase $\text{IrO}_{\text{3}}$ is likely formed and becomes evaporated, posing extreme difficulties for \textit{in}-\textit{situ} growth. We initially deposited stoichiometric  $\text{Nd}_{\text{2}}$$\text{Ir}_{\text{2}}$$\text{O}_{\text{7}}$ films on amorphous phase by growing at $\text{600 }^{\text{o}}$C. Then, we post-annealed the films in air in an electrical box furnace at $\text{1,000 }^{\text{o}}$C for an hour [22].

\begin{figure}
	\includegraphics{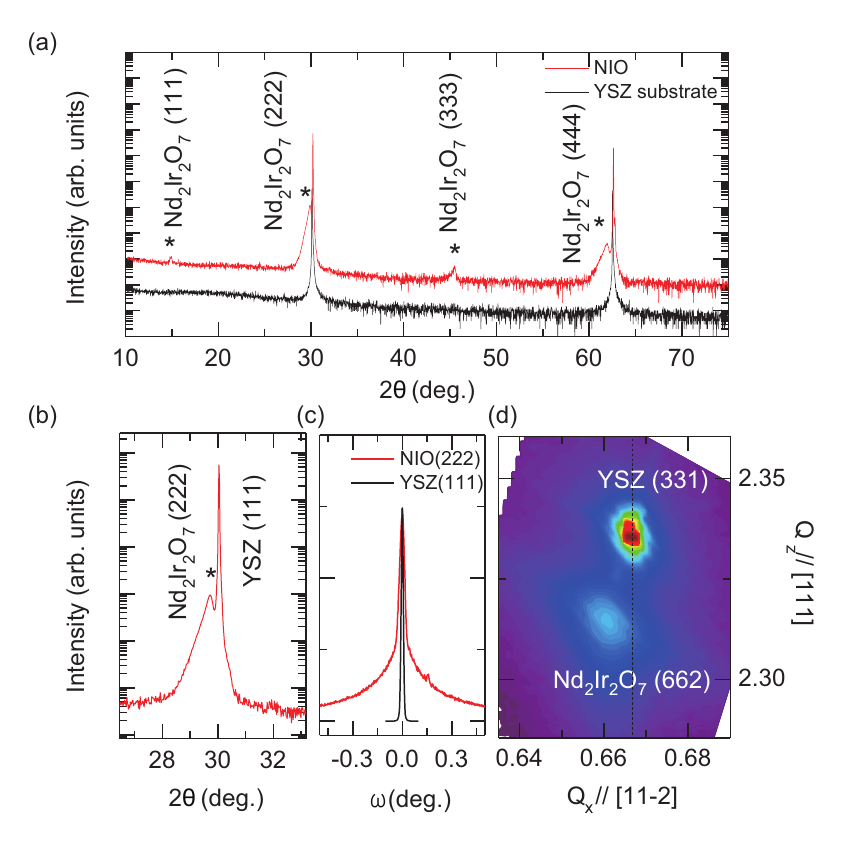}
	\caption{\label{fig1}(Color online) (a) X-ray diffraction pattern of an 80-nm-thick $\text{Nd}_{\text{2}}$$\text{Ir}_{\text{2}}$$\text{O}_{\text{7}}$ film grown on a YSZ (111) substrate (red bold line) and the diffraction pattern of the bare YSZ substrate (black bold line). (b) $\text{Nd}_{\text{2}}$$\text{Ir}_{\text{2}}$$\text{O}_{\text{7}}$ (222) diffraction pattern. (c) Rocking curve of $\text{Nd}_{\text{2}}$$\text{Ir}_{\text{2}}$$\text{O}_{\text{7}}$ /YSZ (111) film. (d) Reciprocal space mapping of $\text{Nd}_{\text{2}}$$\text{Ir}_{\text{2}}$$\text{O}_{\text{7}}$ (662).}
\end{figure}

Figure 2(a) shows X-ray diffraction (XRD) $\theta$-2$\theta$ scans of our $\text{Nd}_{\text{2}}$$\text{Ir}_{\text{2}}$$\text{O}_{\text{7}}$ film and the YSZ substrate. The XRD of film shows only $\text{Nd}_{\text{2}}$$\text{Ir}_{\text{2}}$$\text{O}_{\text{7}}$ (\textit{lll}) peaks in addition to substrate peaks, indicating that the film is single-phase. Figure 2(b) shows the detailed XRD pattern near the   $\text{Nd}_{\text{2}}$$\text{Ir}_{\text{2}}$$\text{O}_{\text{7}}$ (222) peak. Figure 2(c) shows that the full-width-half-maximum (FWHM) of the rocking curve is ~$\text{0.05}^{\text{o}}$, indicating that the sample is highly crystalline. We used X-ray reciprocal space mapping (X-RSM) to measure the in-plane lattice constants of the film and substrate. Figure 2(d) shows X-RSM data around the (662) and (331) Bragg reflections of the $\text{Nd}_{\text{2}}$$\text{Ir}_{\text{2}}$$\text{O}_{\text{7}}$ film and YSZ substrate, respectively. The (662) Bragg peak of the $\text{Nd}_{\text{2}}$$\text{Ir}_{\text{2}}$$\text{O}_{\text{7}}$ film has smaller $\textit{Q}_{\text{x}}$- and $\textit{Q}_{\text{z}}$-values than those of the YSZ (331) Bragg peak. From the experimental $\textit{Q}_{\text{x}}$- and $\textit{Q}_{\text{z}}$-values, the lattice constants of our film are estimated to be \textit{a} = \textit{b} = 10.380 $\buildrel _\circ \over {\mathrm{A}}$, close to those of bulk polycrystalline $\text{Nd}_{\text{2}}$$\text{Ir}_{\text{2}}$$\text{O}_{\text{7}}$ (\textit{a} = \textit{b} = \textit{c} =10.375 $\buildrel _\circ \over {\mathrm{A}}$). This indicates that our films have high structural qualities close to the bulk counterparts.

\subsection{B. Scanning transmission electron microscopy (STEM) measurements}

We used STEM to explore the microstructure of our $\text{Nd}_{\text{2}}$$\text{Ir}_{\text{2}}$$\text{O}_{\text{7}}$ films. Figure 3(a) shows a STEM image of an $\text{Nd}_{\text{2}}$$\text{Ir}_{\text{2}}$$\text{O}_{\text{7}}$ film grown on YSZ (111) with the zone axis parallel to [2$\bar{1}$$\bar{1}$]. A blurry region is evident near the interface between film and substrate, which may be attributable to defects and/or electron beam damage during STEM measurements. To determine structural relationships, we performed fast Fourier transforms (FFTs) of real-space images of the $\text{Nd}_{\text{2}}$$\text{Ir}_{\text{2}}$$\text{O}_{\text{7}}$ film and the YSZ substrate region [see Figs. 3(b) and (c)]. Although our films were grown by post-annealing the amorphous phase, an epitaxial relationship is evident between $\text{Nd}_{\text{2}}$$\text{Ir}_{\text{2}}$$\text{O}_{\text{7}}$ and the YSZ substrate: $\text{Nd}_{\text{2}}$$\text{Ir}_{\text{2}}$$\text{O}_{\text{7}}$ [111]//YSZ [111] and $\text{Nd}_{\text{2}}$$\text{Ir}_{\text{2}}$$\text{O}_{\text{7}}$ [0$\bar{1}$1]//YSZ [0$\bar{1}$1].

\begin{figure}
\includegraphics{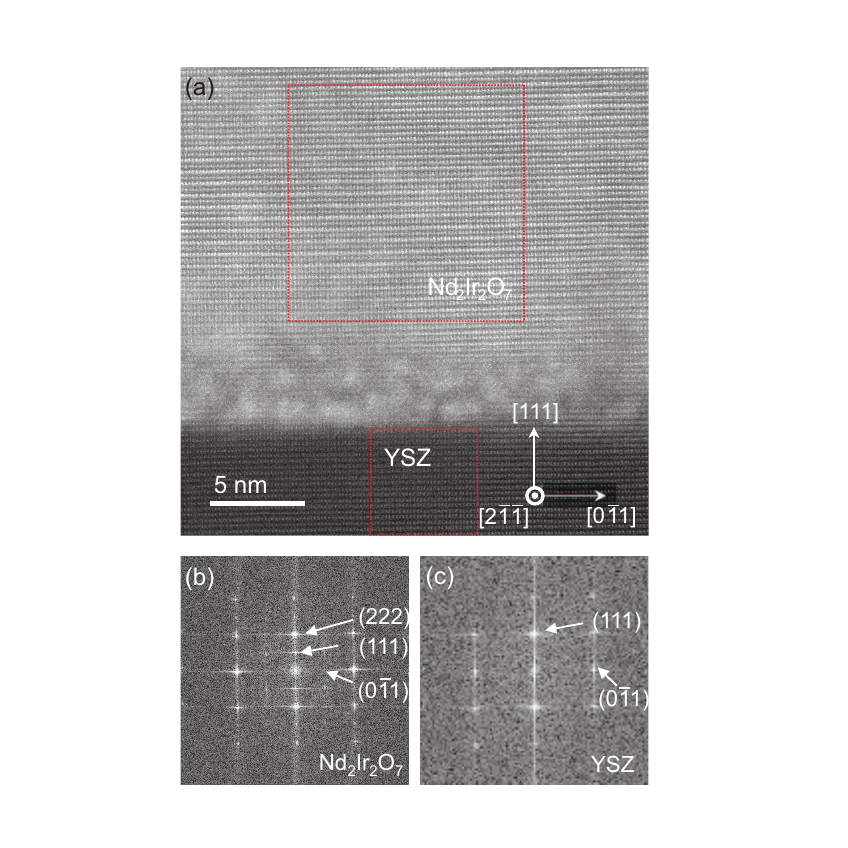}
\caption{\label{fig1}(Color online) (a) Scanning transmission electron microscopy (STEM) image of an $\text{Nd}_{\text{2}}$$\text{Ir}_{\text{2}}$$\text{O}_{\text{7}}$ film grown on YSZ (111) from the viewpoint of the [2$\bar{1}$$\bar{1}$] axis. The $\text{Nd}_{\text{2}}$$\text{Ir}_{\text{2}}$$\text{O}_{\text{7}}$ film exhibits an ordered pyrochlore structure and an epitaxial relationship with the YSZ substrate. (b) The FFT of the pyrochlore $\text{Nd}_{\text{2}}$$\text{Ir}_{\text{2}}$$\text{O}_{\text{7}}$ thin film (upper red dashed square) and (c) that of the YSZ substrate (downside red dashed square).}
\end{figure}

\subsection{C. Resistivity and magneto-transport measurements}

We measured $\textit{dc}$ magneto-transport properties below $\sim$14 T using the standard four-probe method and a commercial cryostat system (Oxford Instruments). For higher $\textit{H}$-fields up to 30 T, we used the resistive magnet at the National High Magnetic Field Laboratory (NHMFL). During these magneto-transport measurements, we applied current along the [1$\bar{1}$0] direction and $\textit{H}$ along the [111] direction (i.e., perpendicular to the current direction). As shown in Fig. 4(a), our 80-nm-thick $\text{Nd}_{\text{2}}$$\text{Ir}_{\text{2}}$$\text{O}_{\text{7}}$ film exhibits a MIT around $\textit{T}_{\text{N}}^{\text{Ir}}$ $\sim$30 K, close to the bulk value. On the other hand, the associated resistivity ratio, $\rho$ (2 K)/$\rho$ (300 K), is much broader than that of the ‘best’ single-crystalline sample [13]. The resistivity ratio, $\rho$ (2 K)/$\rho$ (300 K), is $\sim$10, which is much smaller than the compared value ($\sim$1,000) of the single crystal. However, it should be noted that the reported resistivity ratio values in literature varies in a wide range of 10 $\sim$ 1,000 even for single crystals [23]. Therefore, our $\text{Nd}_{\text{2}}$$\text{Ir}_{\text{2}}$$\text{O}_{\text{7}}$ films would have crystal quality at least comparable to some single crystals [23].

To measure magneto-resistance ($\textit{MR}$) behavior, we zero-field-cooled the $\text{Nd}_{\text{2}}$$\text{Ir}_{\text{2}}$$\text{O}_{\text{7}}$ thin film to 2 K and measured resistivity under a constant $\textit{H}$-field while warming to 300 K. As shown in Fig. 4(b), the resistance changes under an $\textit{H}$-field are large but become smaller as \textit{T} increases, vanishing around $\textit{T}_{\text{N}}^{\text{Ir}}$. Figure 4(c) shows \textit{T}-dependent normalized $\textit{MR}$ curves at various $\textit{H}$ values. The $\textit{MR}$ peaks at $\sim$15 K, which corresponds to the N$\acute{e}$el ordering temperature of the Nd moment ($\textit{T}_{\text{N}}^{\text{Nd}}$) [24].

\begin{figure}
\includegraphics{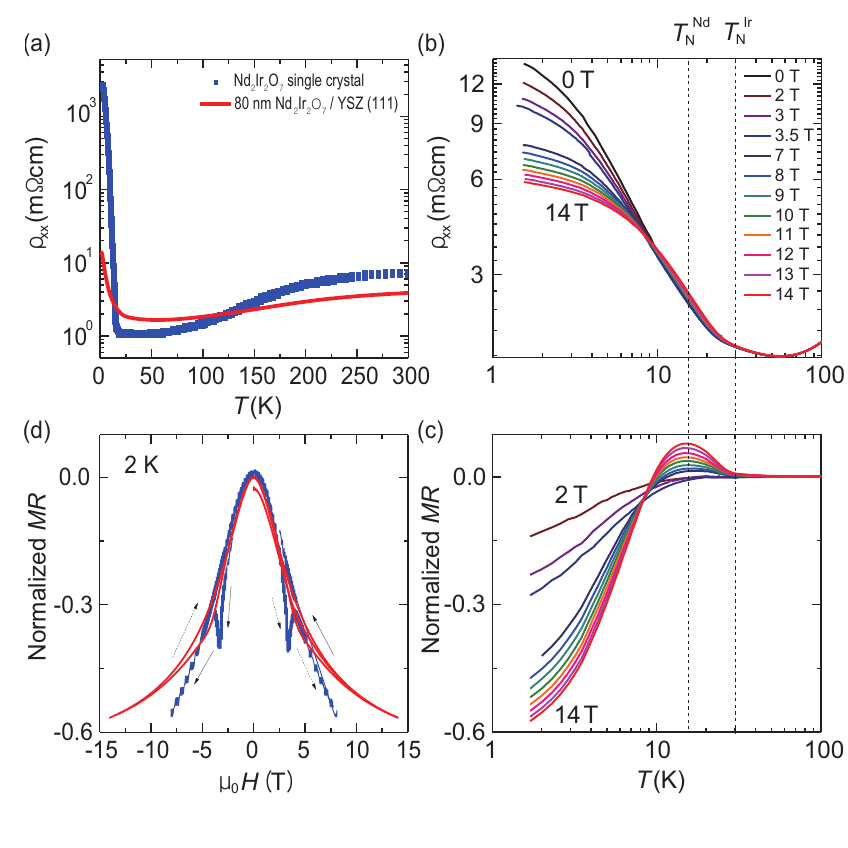}
\caption{\label{fig1}(Color online) (a) Temperature dependence of the longitudinal resistivity $\rho_{xx}$ of the 80-nm-thick film (red bold line) and its counterpart of single crystal from Ref. [13] (blue square). (b) Temperature dependent magneto-resistance ($MR$) of $\text{Nd}_{\text{2}}$$\text{Ir}_{\text{2}}$$\text{O}_{\text{7}}$ for various $H$ // [111] from 0 T to 14 T. The $\textit{T}_{\text{N}}^{\text{Nd}}$ ($\sim$15 K) and $\textit{T}_{\text{N}}^{\text{Ir}}$ ($\sim$30 K) indicate the Néel temperature of the Nd and Ir moments, respectively. (c) Corresponding normalized MR values versus temperature. When H is sufficiently large, the gradient of $MR$ changes from positive to negative at $\textit{T}_{\text{N}}^{\text{Nd}}$. (d) Normalized $MR$ values with $H$ // [111] at 2 K of 80-nm-thick film (red bold line) and a single crystal from Ref. [13] (blue square).}
\end{figure}

\section*{III. ANTIFERROMAGNETIC DOMAIN SWITCHING}
\subsection{A. Magneto-resistance hysteresis}

Figure 4(d) shows $\textit{H}$-dependent normalized $\textit{MR}$ curve at 2 K, where the comparison with the single crystal data in Ref. [13] can be made. Note that it exhibits an intriguing hysteretic behavior with very broad $\textit{MR}$ dips around $\pm$3 T. This feature is somewhat broad, since it might come from many inherent defects inside the film and pining of DWs to such defects. However, this hysteretic behavior is essentially similar to that of single crystal, marked by the blue square line. The $\textit{MR}$ dips occurs around $\pm$3 T, nearly at the magnetic coercive field $\textit{H}_{\text{C}}$ of the single crystal [13]. In addition, the hysteresis direction is the same as that in the single crystal data, suggesting that the magnetic hysteresis might come from DW switching.

Such hysteretic behavior is also evident at other temperatures below $\textit{T}_{\text{N}}^{\text{Ir}}$. Figure 5(a) shows normalized \textit{MR} curves with \textit{H} along the [111] direction at various temperatures. The \textit{MR} is positive at higher \textit{T} but becomes negative around $\textit{T}_{\text{N}}^{\text{Nd}}$ $\sim$15 K [24], consistent with Fig. 4(c). Note that the dip structures in \textit{MR} hysteresis curve can be easily observed between 5 K and 15 K. And these structures occur at the magnetic fields which are close to the reported $\textit{H}_{\text{C}}$ values of single crystals [13]. In the earlier work, the hysteretic \textit{MR} behaviors were already attributable to AFM domain-switching [25]. Likewise, we should be able to explain our \textit{MR} data in terms of switching between the AIAO and AOAI domains.

\begin{figure}
\includegraphics{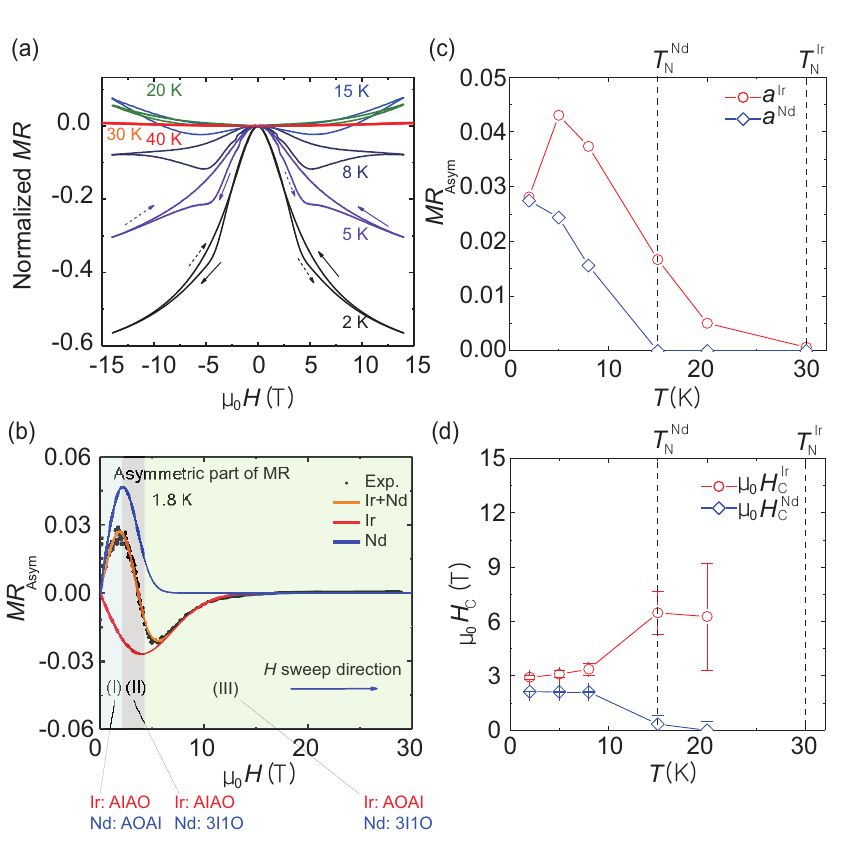}
\caption{\label{fig1}(Color online) (a) Normalized magneto-resistance ($MR$) at various temperatures with $H$ // [111]. (b) Asymmetric part of sweep-up $MR$ at 1.8 K (positive $H$). The colored region with Roman numerals corresponds to the represented spin structures. The red (blue) and orange solid lines indicate the hysteretic part of Ir (Nd) domain switching from Eq. (2) and its total contribution to both Nd and Ir, respectively. (c) Quantified fit parameters '\textit{a}' extracted from Eq. (2), where red empty circles (blue empty diamonds) indicate the contribution of hysteretic behavior from Ir (Nd) domain switching. (d) Red empty circles and blue empty diamonds are the coercive fields of the Ir domain and Nd spins, respectively, from $MR$. At 30 K, the asymmetric $MR$ becomes too small to obtain reliable values of the coercive field.}
\end{figure}

\subsection{B. Asymmetric analysis on hysteretic magneto-resistance curves}

In the hysteresis curve of a ferroic material, it is well known that the asymmetric \textit{MR} part contains important information on magnetic domain switching [22, 26]. We extract the asymmetric part of the normalized \textit{MR} curves using

\begin{equation}
MR_{Asym}(H)=\frac{\rho_{xx}(H)-\rho_{xx}(-H)}{2\rho_{xx}(0)}. \label{pauli}
\end{equation}

Figure 5(b) shows the $\textit{MR}_{\textit{Asym}}(H)$ curve at 1.8 K with increasing \textit{H}-field from $-$30 to 30 T [only the positive side is plotted, given that $MR_{Asym}(-H) = -MR_{Asym}(H)$]. Note that the $\textit{MR}_{\textit{Asym}}(H)$ curve had interesting structures: a peak at lower \textit{H} and a dip at higher \textit{H}. Additionally, the curve displayed a sign change in $\textit{MR}_{\textit{Asym}}(H)$ around 3.0 T. Up to our knowledge, these $\textit{MR}_{\textit{Asym}}(H)$ behaviors have not been observed in other magnetic ferroic materials.

Using detailed analysis of the \textit{MR} curves, we can obtain insights into spin-ordering and DW dynamics. It is well established in most ferroic materials that domain switching usually occurs at $\textit{H}_{\text{C}}$. When a switching becomes broadened due to pinning of DWs to defects and/or surfaces, the $\textit{H}_{\text{C}}$ distribution in DW switching dynamics can be fitted with a Gaussian function [27, 28]. To account for the asymmetry, we fitted our $\textit{MR}_{\textit{Asym}}(H)$ data with an asymmetric Gaussian function:

\begin{equation}
MR_{Asym,Fit}(H)=a(e^{-(H-H_c)^2}-e^{-(H+H_c)^2}), \label{pauli}
\end{equation}

where '\textit{a}' and '\textit{b}' are the magnitude and width of the hysteretic behavior, respectively. 
We found that the associated sign change in $\textit{MR}_{\textit{Asym}}(H)$ in Fig. 5(b) cannot be fit with a single asymmetric Gaussian function in Eq. (2). This indicates that the hysteresis cannot be explained in terms of magnetic switching in one kind of sublattice, i.e., either Ir or Nd spins. On the other hand, when we introduce two magnetic switching functions (namely the two asymmetric Gaussian functions), we are able to fit the experimental $\textit{MR}_{\textit{Asym}}(H)$ curves quite well. In Fig. 5(b), we plot the red (blue) line for higher (lower) field magnetic switching. These indicate that both Ir and Nd spins are involved with magneto-transport properties of our $\text{Nd}_{\text{2}}$$\text{Ir}_{\text{2}}$$\text{O}_{\text{7}}$ films.

Considering the $\textit{T}_{\text{N}}$ of each ion, we find that the switching at the higher (lower) $\textit{H}_{\text{C}}$-value is a result of the Ir (Nd) sublattice. Figure 5(c) shows the temperature dependence of each respective ion’s fit coefficient ('\textit{a}') calculated from the symmetry analysis of $\textit{MR}_{\textit{Asym}}(H)$ curves. As the temperature decreased, the '\textit{a}' value of the red line first appeared around 30 K ($\sim$$\textit{T}_{\text{N}}^{\text{Ir}}$). This coincides with Ir spin ordering; thus, we can assign the red line to $\textit{a}^{\text{Ir}}$. Likewise, the blue line began to emerge around $\textit{T}_{\text{N}}^{\text{Nd}}$ and monotonically increased, thus corresponding to $\textit{a}^{\text{Nd}}$. Note that the values of $\textit{a}^{\text{Ir}}$ are larger than those of $\textit{a}^{\text{Nd}}$. As the conduction in $\text{Nd}_{\text{2}}$$\text{Ir}_{\text{2}}$$\text{O}_{\text{7}}$occurs in the Ir-O network [9], it is reasonable to expect Ir spin to dominate the change in \textit{MR}. Therefore, the red and blue lines in Fig. 5(b) should correspond to switching of spins of Ir and Nd sublattices, respectively.

\subsection{C. Domain switching caused by the \textit{f}-\textit{d} exchange interaction}

Figure 5(b) shows that the switching of Nd spins occurred at lower $H$ than that of Ir spins. Based on the relative values of the switching $H$-field, we can summarize the domain switching sequence as follows: (I) at small $H$, the Ir-AIAO domains dominate, along with Nd-AOAI spin structures; (II) as $H$ increases, Nd spins flip first to form Nd-3I1O while the Ir sublattice remains in the Ir-AIAO configuration; and (III) finally, at higher H, the Nd-3I1O spin structure flips the Ir spins, resulting in the formation of Ir-AOAI domains. This switching sequence is consistent with that was described in Fig. 1.

The \textit{MR} data also indicate that Ir and Nd spins are strongly coupled, presumably due to the \textit{f}-\textit{d} exchange interaction. From the \textit{T}-dependent analysis of $\textit{MR}_{\textit{Asym}}(H)$ we were able to determine the coercive fields required to switch the spins of each Ir and Nd sublattice. The fitted values for $\mu_{0}H_{\text{C}}^{\text{Nd}}$ are shown as blue open diamonds in Fig. 5(d). As expected, Nd spin switching did not occur for \textit{T} \textgreater $\textit{T}_{\text{N}}^{\text{Nd}}$. Below  $\textit{T}_{\text{N}}^{\text{Nd}}$, $\mu_{0}H_{\text{C}}^{\text{Nd}}$ increased and remained nearly constant around 2 T. The fitted values for $\mu_{0}H_{\text{C}}^{\text{Ir}}$ are shown as red open circles. Ir-AIAO to AOAI domain switching began below $\textit{T}_{\text{N}}^{\text{Ir}}$, i.e., $\mu_{0}H_{\text{C}}^{\text{Ir}}$ $\sim$7 T around 20 K. The \textit{MR} signal became small close to $\textit{T}_{\text{N}}^{\text{Ir}}$, making it difficult to estimate a value for $\mu_{0}H_{\text{C}}^{\text{Ir}}$. As the temperature decreased, $\mu_{0}H_{\text{C}}^{\text{Ir}}$ decreased abruptly below $\textit{T}_{\text{N}}^{\text{Nd}}$ and then remained nearly constant around 3 T. Note that the extraordinary decrease of $\mu_{0}H_{\text{C}}^{\text{Ir}}$ around $\textit{T}_{\text{N}}^{\text{Nd}}$ cannot occur in magnetic systems with a single magnetic sublattice. This surprising behavior emphasizes the important role of the \textit{f}-\textit{d} exchange interaction in Ir domain switching of the pyrochlore iridate film.

\section*{IV. ANOMALOUS HALL EFFECTS CAUSED BY ANTIFERROMAGNETIC DOMAIN WALLS}
\subsection{A. AHEs observed in $\text{Nd}_{\text{2}}$$\text{Ir}_{\text{2}}$$\text{O}_{\text{7}}$ films}

We measured the transverse magneto-resistance, i.e., Hall resistivity $\rho_{xy}$, of our $\text{Nd}_{\text{2}}$$\text{Ir}_{\text{2}}$$\text{O}_{\text{7}}$ films with $H$ along the [111] direction. Figure 6(a) shows that the $\rho_{xy}$ curves have a very large and unconventional Hall resistivity behavior. As the temperature decreased below 30 K, hysteretic behavior with a ‘humplike’ peak was observed, which cannot be explained by the conventional ordinary Hall effect. This hysteretic behavior with ‘humplike’ peak intensity became more pronounced as the temperature decreased below $\textit{T}_{\text{N}}^{\text{Nd}}$. Similar ‘humplike’ behavior has been observed in ferromagnetic DWs and has been attributed to the presence of a real-space topological Hall effect related to magnetic skyrmions [5, 6]. 

To obtain further insight into the AHE, we subtracted the $H$-linear ordinary Hall effect term from the 2 K $\rho_{xy}$ data and plotted the data as black solid circles in Fig. 6(b). As shown in the figure, the AHE term becomes peaked around 2 $\sim$3 T. The corresponding $H$ value is close to $\mu_{0}H_{\text{C}}^{\text{Ir}}$ values, obtained from the earlier $MR$ data analysis. It indicates that a DW could play a significant role in the observed AHE. In addition, the experimental AHE reached a nearly constant value above 5 T, suggesting that bulk effects may also be involved. Taken together, these results suggest that the AHE may have originated from both bulk and DW contributions, hereafter referred to as bulk-AHE and DW-AHE, respectively.

\begin{figure}
\includegraphics{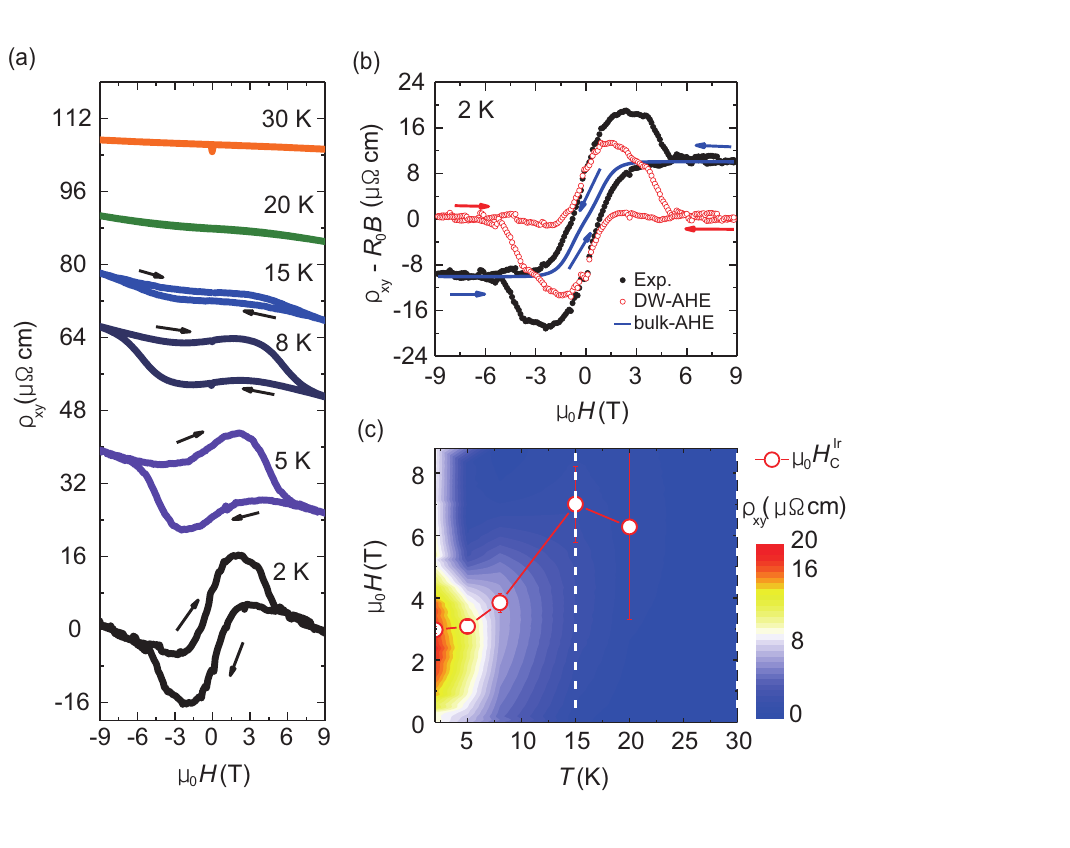}
\caption{\label{fig1}(Color online) (a) $H$-dependence of Hall resistivity taken with $H$ // [111] at various temperatures. The arrows indicate the $H$ sweep directions. (b) Black empty circles indicate the Hall resistivity at 2 K. The ordinary Hall term is subtracted by linear fitting in the higher $H$ region. The blue solid lines indicate the contribution of the bulk-anomalous Hall effect (AHE). The red circles indicate the DW-AHE. (c) Color map of anomalous Hall resistivity (ordinary Hall term is subtracted) in the $T$-$H$ plane. The red empty circles represent the coercive field $H_{C}^{Ir}$ of Ir domain switching, which are obtained from $MR$ measurements.}
\end{figure}

\subsection{B. Bulk-AHE of antiferromagnetic material: Scalar spin chirality}

For AFM bulk materials, AHEs have been observed and explained in terms of the scalar spin chirality. This term can be defined as $\vec{S}_{i}$$\cdot$($\vec{S}_{j}$$\times$$\vec{S}_{k}$) with three local spins $\vec{S}_{i}$, $\vec{S}_{j}$ and $\vec{S}_{k}$ [29]; the scalar product corresponds to the solid angle $\Omega$ subtended by the three spins. The noncoplanar spin configuration is closely linked to the Berry curvature and can generate a fictitious magnetic flux proportional to the scalar spin chirality. As a result, AHE can occur [30-37]. Without a magnetic field, the total magnetization and scalar spin chirality remain zero inside a single $\text{Nd}_{\text{2}}$$\text{Ir}_{\text{2}}$$\text{O}_{\text{7}}$ magnetic domain. However, applying the strong magnetic field changes the order of the Nd moments to 3I1O [see Fig. 1(b)] and Ir moments to a canted AIAO. Then, both magnetization and scalar spin chirality become finite, and the canted AFM exhibits finite bulk-AHE.

Now let us estimate $\rho_{xy}$ values attributable to the bulk-AHE for $\text{Nd}_{\text{2}}$$\text{Ir}_{\text{2}}$$\text{O}_{\text{7}}$. Note that the AIAO (AOAI) order of the Ir moments is canted by the 1I3O (3I1O) order of the Nd moments via the $f$-$d$ exchange interaction. As a result, the scalar spin chirality is produced in the Ir spin system and generates a fictitious magnetic field parallel to its magnetization direction [38]. Moreover, it is well-known that the Nd moments ($\sim$2.4 $\mu_{B}$/Nd) are much larger than the Ir moments ($\sim$0.2 $\mu_{B}$/Ir) [13]. As the magnetic moment of Ir is an order of magnitude smaller than that of Nd, the Ir sublattice makes a much smaller contribution to the magnetization. Given that both the scalar spin chirality term of Ir spins and total magnetization are dominated by Nd moments, we will focus on the spin configuration in Nd tetrahedrons, changing from AIAO to 3I1O. We can quantify the proportions of Nd unit cells in the 3I1O order by calculating the expectation value of magnetization per unit cell. Then we can infer a bulk Hall resistivity curve, which is proportional to the number of Nd unit cells in the 3I1O order.

To quantitatively estimate the $H$-dependent net magnetization of the Nd sublattice, we consider the Hamiltonian below, describes a single Nd unit cell under a magnetic field in the direction of the [111] plane:

\begin{equation}
H=-J\sum_{<ij>}\vec{S}_{i}\cdot\vec{S}_{j}-\mu\vec{B}\cdot \sum_{i}\vec{S}_{i}-K\Phi B_xB_yB_z, \label{pauli}
\end{equation}

where $J$ is the interaction strength, $\mu$ is the Bohr magneton, and $\vec{B}$ is magnetic field. The $K$ is a coefficient and $\Phi$ is the AIAO order parameter expressed as

\begin{equation}
\begin{aligned}
\Phi={} & \frac{1}{4\sqrt{3}}(S_{1x}+S_{1y}+S_{1z}+S_{2x}-S_{2y}-S_{2z}\\
& -S_{3x}+S_{3y}-S_{3z}-S_{4x}-S_{4y}+S_{4z}). \label{pauli}
\end{aligned}
\end{equation}

Here, we treat the Nd spin as an Ising variable. The spin vector at each site in a unit cell is defined as $\vec{S}_{1}$ = $\frac{\sigma_{1}}{\sqrt{3}}$(1,1,1), $\vec{S}_{2}$ = $\frac{\sigma_{2}}{\sqrt{3}}$(1,-1,-1), $\vec{S}_{3}$ = $\frac{\sigma_{3}}{\sqrt{3}}$(-1,1,-1) and $\vec{S}_{4}$ = $\frac{\sigma_{4}}{\sqrt{3}}$(-1,-1,1), where $\sigma_{i}$ = $\pm 1$. Then, the Eq. (3) can be simplified to

\begin{equation}
H=-J\sum_{<ij>}\sigma_{i} \sigma_{j}-\mu\vec{B}\cdot \sum_{i}\vec{S}_{i}-K\Phi B_xB_yB_z, \label{pauli}
\end{equation}

When $J$ \textgreater 0, the ground state of the Hamiltonian in a Nd unit cell is AIAO or AOAI order without the magnetic field; namely, $\sigma_{i}$ are either +1 or -1. As such, the expectation value of the magnetization can be calculated using a partition function. For the unit cell, the partition function is

\begin{equation}
Z=\sum_{\sigma_{1},\sigma_{2},\sigma_{3},\sigma_{4}}exp(-\beta H), \label{pauli}
\end{equation}

and the expectation value of magnetization along [111] direction is

\begin{equation}
<M>=\frac{1}{Z} \sum_{\sigma_{1},\sigma_{2},\sigma_{3},\sigma_{4}}(\sigma_{1}-\frac{1}{3} (\sigma_{2}+\sigma_{3}+\sigma_{4}))exp(-\beta H), \label{pauli}
\end{equation}

We evaluated the bulk-AHE of our $\text{Nd}_{\text{2}}$$\text{Ir}_{\text{2}}$$\text{O}_{\text{7}}$ film by assuming that it should be proportional to the magnetization. Given a sufficient $H$-field, all of the DWs should disappear, such that the DW-AHE becomes negligible and the bulk-AHE dominates. Using the experimental values of the saturated Hall resistivity value ($\sim$8 $\mu\Omega$ cm) at $\sim$2 T ($\mu_{0}H_{\text{C}}^{\text{Nd}}$) in Fig. 6(b), we derived an analytical estimate of the bulk-AHE contribution. The bulk-AHE contribution is shown as the blue line in Fig. 6(b). It should be noted that the bulk-AHE alone cannot explain the experimental AHE. There should an additional AHE, which cannot be explained by the scalar spin chirality effect.

\begin{figure}
\includegraphics{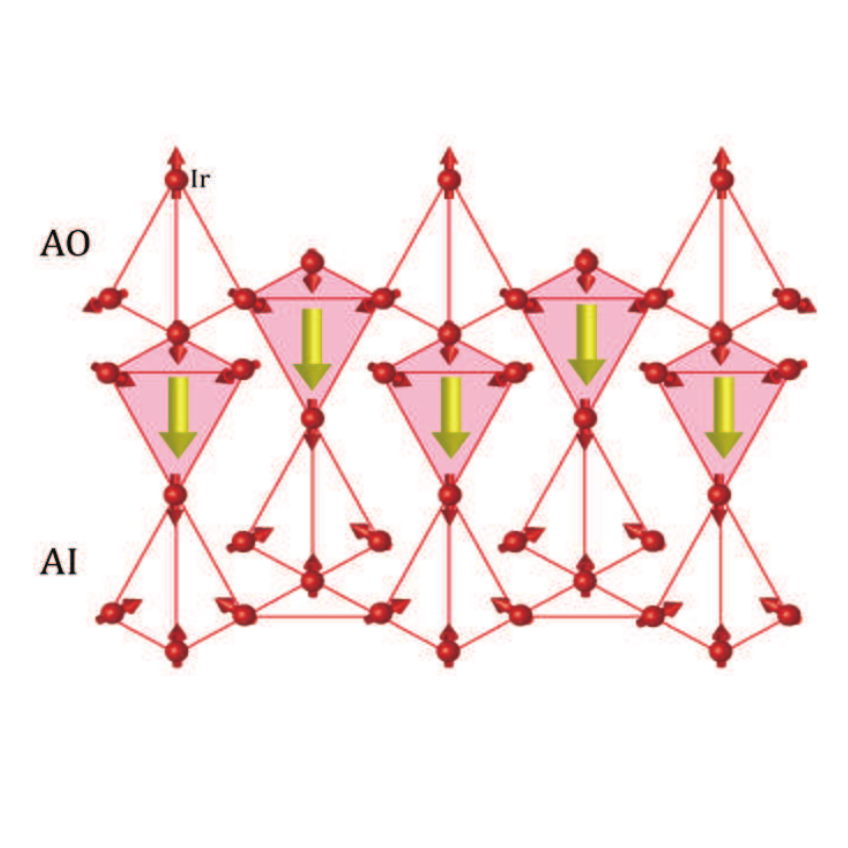}
\caption{\label{fig1}(Color online) As an iridium domain wall (DW) in the bulk breaks the two-fold rotation symmetry in pyrochlore iridates, anomalous Hall conductivity can be nontrivial. Due to the remaining symmetries with a DW, finite anomalous Hall conductivity can arise on the plane parallel to the DW.}
\end{figure}

\subsection{C. Emergence of the AHE in the presence of an AFM domain wall: Symmetry analysis}

We considered the possible emergence of the AHE due to momentum space Berry curvature by performing symmetry analysis. First, we should note that the anomalous Hall conductivity cannot occur in a single domain of AIAO order in cubic pyrochlore iridates [15]. To calculate the intrinsic anomalous Hall conductivity, we integrate the Berry curvature of the occupied bands in momentum space:

\begin{equation}
\sigma_{\alpha \beta}=\frac{e^{2}}{\hbar}\int_{BZ} \frac{d^{3}k}{(2 \pi)^{3}} \sum_{n} f(\epsilon_{n}(\vec{k})-\mu)F_{\alpha \beta} (\vec{k}), \label{pauli}
\end{equation}

where $\textit{f}$ ($\epsilon_{n}$($\vec{k}$)-$\mu$) is the Fermi-Dirac distribution, and  $\textit{F}_{\alpha \beta}$ is the Berry curvature of $\alpha \beta$ plane. For convenience, we denote each component of the Berry curvature in a vector notation, i.e., $\textit{F}^{x}$($\vec{k}$) = $\textit{F}_{yz}$($\vec{k}$), $\textit{F}^{y}$($\vec{k}$) = $\textit{F}_{zx}$($\vec{k}$) and $\textit{F}^{z}$($\vec{k}$) = $\textit{F}_{xy}$($\vec{k}$). When two-fold rotation symmetry around the z-axis ($C_{2z}$) exists, as in the single domain of AIAO (or AOAI) order, $\textit{F}^{x}$(-$\textit{k}_{x}$, -$\textit{k}_{y}$, $\textit{k}_{z}$) = $-F^{x}(k_{x}, k_{y}, k_{z})$, $F^{y}(-k_{x}, -k_{y}, k_{z})$ = $-F^{y}(k_{x}, k_{y}, k_{z})$, and $F^{z}(-k_{x}, -k_{y}, k_{z})$ = $F^{z}(k_{x}, k_{y}, k_{z})$. Then $\sigma_{yz}$,$\sigma_{zx}$ should be trivial, but $\sigma_{xy}$ does not have to be trivial through Eq. (8). In a single domain of AIAO order, cubic pyrochlore iridates have $C_{2x}$, $C_{2y}$ and $C_{2z}$ symmetry, thus, AHEs cannot occur.

However, in the presence of a DW, an unconventional AHE can be generated from the non-zero Berry curvature. At the DW, two-fold rotation symmetries become broken but there exists a three-fold rotational symmetry about the axis perpendicular to the DW [see Fig. 7]. Considering the three-fold rotation axis as [111] and taking the two other perpendicular axes as [0$\bar{1}$1] and [$\bar{2}$11], the symmetry properties of the Berry curvature under three-fold rotational symmetry are $F^{x}(k_{y}, k_{z}, k_{x})$ = $F^{y}(k_{x}, k_{y}, k_{z})$, $F^{y}(k_{y}, k_{z}, k_{x})$ = $F^{z}(k_{x}, k_{y}, k_{z})$ and $F^{z}(k_{y}, k_{z}, k_{x})$ = $F^{x}(k_{x}, k_{y}, k_{z})$. Using these properties, we establish the properties of the Berry curvature components along [111] as

\begin{equation}
F^{[111]} (\vec{k})+F^{[111]} (C\vec{k})+F^{[111]} (C^{-1}\vec{k}) \neq 0, \label{pauli}
\end{equation}

where $C$ is the three-fold rotation operator. Therefore, $\sigma^{[111]} \neq 0$, and finite anomalous Hall conductivity can occur in the plane parallel to the DW.

\subsection{D. Unconventional AHE at the antiferromagnetic domain walls of $\text{Nd}_{\text{2}}$$\text{Ir}_{\text{2}}$$\text{O}_{\text{7}}$ films}

When $H$ is applied parallel to the [111] direction, DWs are expected to form within the (111) plane. Among the possible DW orientations (see Appendix A), in a zero $H$-field, the AIAO magnetic ground state favors the formation of DWs within the (111) plane due to its lower frustration compared with other possible orientations [39, 40]. Moreover, a magnetization experiment on $\text{Cd}_{\text{2}}$$\text{Os}_{\text{2}}$$\text{O}_{\text{7}}$ single crystal, another AIAO ordered pyrochlore structure, showed that the plane of DW formation prefers to orient normal to the direction of the applied $H$ [40]. Thus, in our experimental geometry, DWs with the (111) plane orientation should play a significant role in magneto-transport results. 
Figure 6(b) shows the $H$-dependent $\rho_{xy}$ at 2 K, from which we subtract the ordinary term that varies linearly with $H$. As mentioned earlier, we should interpret this unconventional Hall resistivity as a combination of the bulk-AHE and the DW-AHE. The bulk-AHE is shown as the blue curve, estimated by following the procedure in Section IV B. To obtain the Hall resistivity attributable to DW-AHE, we subtracted the bulk-AHE from the experimental values. The red circles in Fig. 6(b) show the remaining contribution.

Note that the maximum value of the remaining Hall resistivity of $\Delta$$\rho_{xy}$  $\sim$15 $\mu\Omega$ cm is twice as large as the saturated Hall resistivity value of $\sim$8 $\mu\Omega$ cm. The obtained maximum value of the red circles ($\sim$2 T) is also in reasonable agreement with $\mu_{0}H_{\text{C}}^{\text{Ir}}$ ($\sim$3 T) from $MR$ in which a DW conductance contribution played a significant role. Moreover, the quadratic dependence of the DW anomalous Hall resistivity on longitudinal resistivity indicates that intrinsic AHEs arose at $\mu_{0}H_{\text{C}}^{\text{Ir}}$ ($\sim$3 T) (see Appendix B). Thus, our results indicate that the ‘humplike’ $\rho_{xy}$ signal originates from the DW-AHE.

	Figure 6(c) shows a contour plot of the $\rho_{xy}$, from which we subtract the ordinary term, as a function of both T and $H$. The contour plot also supports a large enhancement of AHE due to DWs. For a given T, the $H$ value at which $\rho_{xy}$ reaches its maximum is correlated with the value of $\mu_{0}H_{\text{C}}^{\text{Ir}}$ from $MR$. This implies that the large density of DWs near $\mu_{0}H_{\text{C}}^{\text{Ir}}$ can result in a maximized contribution to the AHE. The peak value of the ‘humplike’ Hall resistivity was also highly enhanced below $T_{\text{N}}^{\text{Nd}}$, which supports the idea of strong coupling between Ir and Nd sublattices via the $f$-$d$ exchange interaction. 

It has already been observed experimentally that the DWs of $\text{Nd}_{\text{2}}$$\text{Ir}_{\text{2}}$$\text{O}_{\text{7}}$ have a much higher conductivity than bulk $\text{Nd}_{\text{2}}$$\text{Ir}_{\text{2}}$$\text{O}_{\text{7}}$. Their conductivity is approximately one order of magnitude larger [13, 41]. Hence, the DW-AHE may be significantly enhanced in $\text{Nd}_{\text{2}}$$\text{Ir}_{\text{2}}$$\text{O}_{\text{7}}$ thin films. In comparison, Hall data on $\text{Sm}_{\text{2}}$$\text{Ir}_{\text{2}}$$\text{O}_{\text{7}}$ films, in which the DW conductance is lower than that of the bulk [42], do not reveal any hysteretic AHE, as shown in Fig. 10 in Appendix C. This observation also indirectly supports that the hysteretic hump-like AHE observed in $\text{Nd}_{\text{2}}$$\text{Ir}_{\text{2}}$$\text{O}_{\text{7}}$ film originates from DW conduction.

\section*{V. SUMMARY}

In summary, we observed a large AHE in an AFM $\text{Nd}_{\text{2}}$$\text{Ir}_{\text{2}}$$\text{O}_{\text{7}}$ thin film, which was induced mainly by the DWs. The strong $f$-$d$ exchange interaction effectively lowers the energy barrier for Ir domain switching, such that a small magnetic field can be used to control the Ir-DW. Eventually, effective Ir-domain switching induces an intrinsic AHE from the Berry curvature at the DW, leading to a large DW-AHE, $\sim$15 $\mu\Omega$ cm at 2 K. Given the observation of AHE at DWs, we suggest that $\text{Nd}_{\text{2}}$$\text{Ir}_{\text{2}}$$\text{O}_{\text{7}}$ is a fertile area for the discovery and investigation of new topological phenomena. As the DW-conducting channels of $\text{Nd}_{\text{2}}$$\text{Ir}_{\text{2}}$$\text{O}_{\text{7}}$ are attributable to the fact that the system lies close to the topological Weyl semimetal phase, the observed DW-driven AHE provides additional evidence that the system exhibits novel topological properties

\section*{ACKNOWLEDGMENTS}

We are grateful to E. S. Choi for helpful discussions and assistance with high magnetic field measurements. We wish to thank Y. Kozuka and K. Ueda for helpful discussions. This work was supported by the Research Center Program of the Institute for Basic Science (IBS) in Korea (IBS-R009-D1 and IBS-R009-G1). The work at the National High Magnetic Field Laboratory was supported by a National Science Foundation (NSF) Cooperative Agreement Grant (No. DMR-1157490) and the State of Florida. T.O. and B.J.Y. acknowledge the support of the Research Resettlement Fund for new faculty of Seoul National University and the Basic Science Research Program through the National Research Foundation of Korea (NRF) funded by the Ministry of Education (Grant No. 0426-20150011). J.H.G. and A.S acknowledge the support of NSF Grant No. DMR-1454200.

\section*{APPENDIX A: ORIENTATION OF MAGNETIC DOMAIN WALL PLANE}

The local structures of DWs are considered in terms of the classical spin model. The stability of DWs has been studied in the extreme limit of the strong Ising anisotropy [40]. Three types of spin configurations are possible at local structures of DWs: Ir-AIAO or Ir-AOAI, Ir-3I1O or Ir-1I3O, and Ir-2I2O. The energies of these configurations are given as $E_{AIAO}$ = −6$J_{eff}$, $E_{3I1O}$ = 0, and $E_{2I2O}$ = 2$J_{eff}$, where $J_{eff}$ is the nearest-neighbor effective antiferromagnetic interaction ($J_{eff}$ {\textgreater} 0). The stability between these DWs can be compared in terms of these energies. The energy of a DW defined as the energy cost per area ($‘A’$) is calculated as $E_{DW}$ = 2($E_{2I2O}$$-E_{AIAO}$)/$A$ = 16 $J_{eff}$ /$A$ for a (100) DW consisting of Ir-2I2O tetrahedra [see Fig. 8(c)]. Similarly, for {\{}111{\}} and {\{}011{\}} DWs, containing only Ir-1I3O tetrahedra as depicted in Fig. 8(a) and (b) are most stable with $E_{DW}$ $\sim$13.9 $J_{eff}$ /$A$ and ~17.0 $J_{eff}$ /$A$, respectively.
Among these possible DWs configurations, {\{}111{\}} DW is the most preferable state due to its lowest $E_{DW}$ value. Other types of DWs also can be formed, and can be understood as the combination of above DWs configurations. Particularly, within the $H$ along the [111], energy cost of (111) plane DW formation should be lower. Therefore, in our experimental geometry, $\text{Nd}_{\text{2}}$$\text{Ir}_{\text{2}}$$\text{O}_{\text{7}}$ film should contain (111) plane DW with $H$ along the [111] direction.

\begin{figure}
\includegraphics{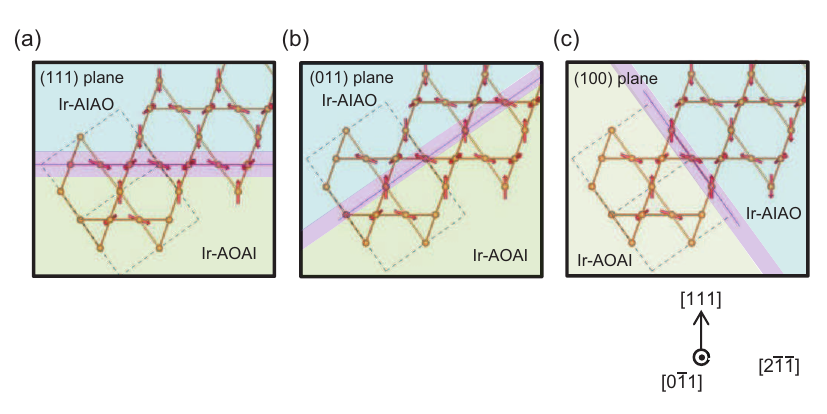}
\caption{\label{fig1}(Color online) Schematic diagrams of various local structures of domain walls (DWs). Blue and light green region indicate the all-in-all-out (AIAO) and all-out-all-in (AOAI) Ir spin ordered domain respectively. The purple region indicate domain wall (DW) between two distinct domains. The dashed line indicates unit cell of $\text{Nd}_{\text{2}}$$\text{Ir}_{\text{2}}$$\text{O}_{\text{7}}$. (a) Spin structures of (111) DW consisting of 1-in-3-out (1I3O) tetrahedra and (b) (011) DW consisting of 1I3O tetrahedra. (c) Spin structures of (100) DW consisting of 2-in-2-out (2I2O) tetrahedra.}
\end{figure}

\begin{figure}[h!]
\includegraphics{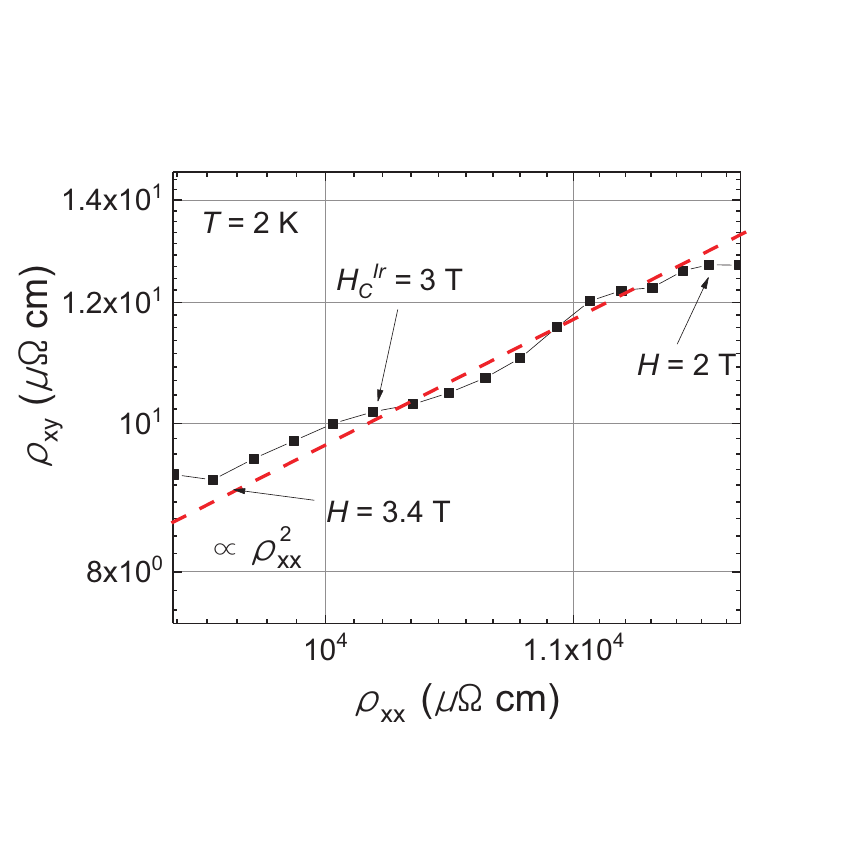}
\caption{\label{fig1}(Color online) The $\rho_{xx}$ versus $\rho_{xy}$ plot at 2 K under the $H$-field along the [111] direction, which is close to coercive field $H_{\text{C}}^{\text{Ir}}$. The dashed lines are to guide the eyes. Note that both $x$-axis and $y$-axis are depicted in the log scale.}
\end{figure}

\section*{APPENDIX B: QUADRATIC DEPENDENCE OF THE DOMAIN WALL ANOMALOUS HALL RESISTIVITY ON LONGITUDINAL RESISTIVITY}

The quadratic dependence of the DW anomalous Hall resistivity on the longitudinal resistivity indicates that the intrinsic AHE arises at $H$ $\sim$ $H_{\text{C}}^{\text{Ir}}$, where a large DW contribution plays a role. Based on the intrinsic mechanism, $\rho_{xy}$ is known to exhibit a power law dependence on $\rho_{xx}$, i.e., $\rho_{xy}$ $\sim$ $\rho_{xx}^{a}$, with the exponent $a$ $\sim$ 2.0 [43]. Near the domain switching region, where the DW contribution is largest, the power law dependence of $\rho_{xy}$ shows $\rho_{xy}$ $\sim$ $\rho_{xx}^{2}$ [see Fig. 9]. Ferromagnetic ordering at the DW may induce the intrinsic AHE, which is usually observed in ferromagnets.

\begin{figure}
\includegraphics{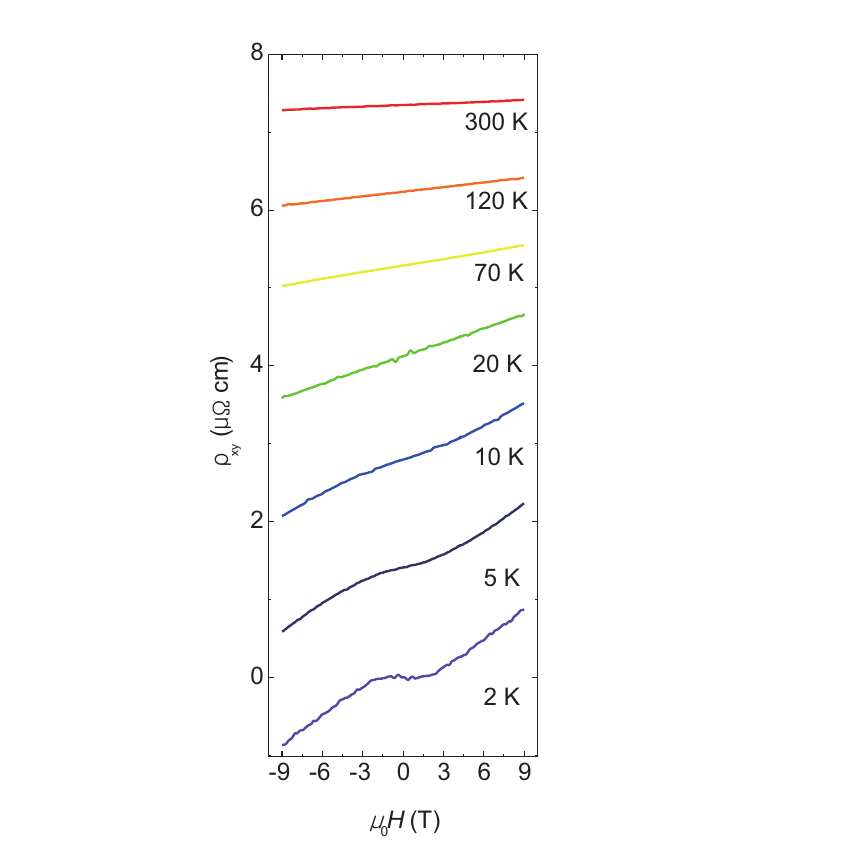}
\caption{\label{fig1}(Color online) $H$-dependence of Hall resistivity of $\sim$170-nm-thick $\text{Sm}_{\text{2}}$$\text{Ir}_{\text{2}}$$\text{O}_{\text{7}}$ film taken with $H$ // [111] at various temperatures.}
\end{figure}

\section*{APPENDIX C: HALL MEASUREMENT ON OTHER RARE EARTH PYROCHLORE IRIDATE FILM}

We measured the Hall resistivity of a $\text{Sm}_{\text{2}}$$\text{Ir}_{\text{2}}$$\text{O}_{\text{7}}$/YSZ (111) thin film, which has a thickness of $\sim$170 nm. The current path was along the [1$\bar{1}$0] direction, and $H$ was always oriented perpendicular to the current direction, which is along the [111] direction. Figure 10 shows the $\rho_{xy}$ curves at various $T$, including the value at $\textit{T}_{\text{N}}^{\text{Ir}}$ $\sim$ 120 K of $\text{Sm}_{\text{2}}$$\text{Ir}_{\text{2}}$$\text{O}_{\text{7}}$. At high temperatures (above $\sim$70 K), the Hall resistivity has a linear dependence on $H$, which is typical behavior for the ordinary Hall effect. However, at lower temperatures (below $\sim$20 K), we observe an anomaly that presumably arises due to the Ir moment canting effect. In comparison to the Hall results for the $\text{Nd}_{\text{2}}$$\text{Ir}_{\text{2}}$$\text{O}_{\text{7}}$ film, no hysteresis occurs. We speculate that this different Hall behavior is induced by the different DW conductance.

\email[E-mail address: {$^*$twnoh@snu.ac.kr}

\vfil\eject

\begin{references}
\bibitem{25}Under the $H$-field, both of spin canting and domain switching can induce $MR$. However, the irreversible processes, such as hysteresis, can be explained in terms of the domain switching.
\end{references}
\end{document}